\documentstyle[aps,preprint]{revtex}
\begin{document}

\title{Metallic nanowires: multi-shelled or filled ?}
 
\draft
\tightenlines

\author{ G. Bilalbegovi\'c }

\address{Department of Physics, University of Rijeka, Omladinska 14,
51 000 Rijeka, Croatia}

\date{to be published in Computational Materials Science}

\maketitle

\begin{abstract}

The room temperature structure of aluminum, copper and gold infinite
nanowires 
is studied. The molecular dynamics simulation method and 
the same type of the embedded atom potentials made by Voter and coworkers
are used. 
It was found that multi-shelled and various filled metallic nanowires exist 
depending on the metal and the initial configuration.
The results were compared with previous simulations for gold nanowires
using different type of the potential. 

Pacs: 61.46.+w,68.65.+g,81.05.Ys


\end{abstract}

\section{Introduction}

The studies of nanostructures are important for advances 
in micro-electronics. Cylindrical nanostructures, i.e., nanowires were
prepared
for various materials \cite{Dekker,Iijima,Tenne}.
The most studied are single and multi-walled carbon nanotubes 
\cite{Dekker,Iijima}. 
Nanowires made of metals are also important 
\cite{Pascual,Mendez,Costa,Sorensen,Bratkovsky,Finbow,Mehrez,Barnett,Gulseren,Ercolessi,Goranka,GBSS,Ikeda}.
Interest
for metallic nanowires especially increased after the tip-surface contact 
experiments performed in 
Scanning Tunneling Microscope (STM) \cite{Mendez}. This was followed by 
a discovery of cylindrical metallic nanostructures in 
the table-top contact \cite{Costa}.
Experimental and theoretical results show that the quantized conductance 
exist for these metallic nanocontacts \cite{Ruitenbeek}.
Molecular Dynamics (MD)
simulation method was also used 
for studies of metallic nanowires and nanocontacts
\cite{Pascual,Sorensen,Bratkovsky,Finbow,Mehrez,Barnett,Gulseren,Ercolessi,Goranka,GBSS,Ikeda}.
Structural, mechanical, and vibrational properties were 
investigated.

Carbon nanotubes often exist as 
multi-shelled structures \cite{Dekker,Iijima}.
Multi-walled nanowires were also found for several inorganic layered materials,
such as $WS_2$, $MoS_2$, and $NiCl_2$ \cite{Tenne}.
Electronic shells in metallic nanowires were obtained
in the jellium model calculation
using the local-density-functional based shells correction method \cite{Yannouleas}.
This was recently confirmed by the conductance measurements for
sodium nanowires formed in the contact between two bulk sodium electrodes 
\cite{Yanson}. MD simulations have shown an existence of finite 
\cite{Goranka} and infinite \cite{GBSS} multi-shelled gold nanowires. 
The radius of these gold nanowires was of the order of $1$ $nm$, 
whereas the length
of simulated finite nanowires was from one to several $nm$.
Similar gold nanostructures were observed in the STM 
and transmission electron microscope experiments \cite{Ohnishi,Rubio}.

Copper, aluminum, and gold are technologically important metals. 
They are widely
applied in macroscopic electric circuits, as well as in mesoscopic devices. 
For example, these metals are used for interconnections in the chips.
Because of their importance 
special methods of producing metallic parts in integrated systems
are developed, such as the electroless copper deposition \cite{Li}. 
Recently applied methods for making nanoscale devices are also
the X-ray lithography and the technique of using an atomic force microscope
cantilever for the stencil \cite{Luthi}. These methods are producing 
wires thinner than $0.1 \mu m$. 
Computer simulations of metallic nanowires should help in developing 
new methods for their fabrication. 
The computational methods also provide access to microscopic information 
about physical properties of nanowires.
In this paper the MD simulation study of internal structure of aluminum,
copper,  and gold nanowires has been carried out. The goal is to determine 
whether multi-shelled structures, already found in MD simulations
for finite and infinite gold nanowires \cite{Goranka,GBSS}, exist for
other metals.

\section{Computational Method}

Embedded atom and effective medium potentials are widely used 
in simulation of metals \cite{Daw,Jacobsen}.
They provide a good description of metallic bonding for bulk, surfaces, 
and clusters.
In comparison with the pair potentials, in the embedded atom model 
there is an additional term for the energy 
to embed an atom in the electron density of its neighbours.
Several versions of these potentials are available.
They differ in the fitting procedure and in the
selection of the properties which were taken for the fit. 
In this work the embedded atom potentials for copper, aluminum, and gold
made by A. Voter and coworkers were used \cite{Liu,Voter,Chen}.
The preparation of these potentials included the fitting of the lattice constant,
the bulk modulus, the cohesive energy, the three cubic elastic constants, 
the vacancy formation energy, as well as
the bond length and bond energy 
of the diatomic molecule.

The classical MD simulation method was applied. The temperature control was
realized through velocity rescaling. 
A time step of $2.64 \times 10^{-15}$ s was used for aluminum, 
$4.06 \times 10^{-15}$ s for copper, and $7.14 \times 10^{-15}$ s for gold.
The number of atoms in nanowires was 
between $1032$ and $1152$. The periodic boundary conditions 
were used along the wire axis.  
The simulation started from ideal 
fcc structures at $T=0$ K. The cross-sections were $(111)$ oriented. 
In previous simulation of infinite gold nanowires 
(using the so-called glue model potential)
it was found that this initial orientation produces
better multi-shelled structures than $(110)$ and $(100)$ \cite{GBSS}.
The cylindrical MD boxes included all atoms whose
distance from the axes was less than $1.2$ $nm$.
The same radius produced a pronounced five-shelled structure 
for an infinite 
$(111)$ oriented gold nanowire in the previous simulation \cite{GBSS}. 
Following the annealing and quenching procedure, the nanowires 
were analyzed at $T=300$ K, 
and  after evolution of up to $10^6$ time steps.

\section{Results and discussion}

Figure 1 shows that this aluminum 
nanowire consists of the five coaxial cylindrical walls. 
The core of the nanowire is empty. The cylindrical walls are disordered.
As presented in Fig. 2, the fcc(111) structure of the cross sections
is preserved in the copper
nanowire.
The side view in Fig. 2(b) shows that the perfect layered 
structure was formed. 
The faceted and filled gold nanowire is presented in Fig. 3.
The parallel rows of atoms in the cross section are apparent. 
This should be compared to 
the initial fcc(111) cross section with the hexagonal symmetry.
It was found that these copper and gold nanowires preserve their filled
structures at higher temperatures.
Therefore, the results show that cylindrical shells
appear in the aluminum nanowire. 
However, it is not obvious why shells were not found for 
the copper and gold nanowires prepared in the same way. 
Moreover, previous simulations \cite{Goranka,GBSS}
for gold nanowires modeled by another type of the potential \cite{Furio},
have shown that multi-shelled structures of lasting stability appear.
There was  a possibility that simulations for copper and gold nanowires
which employ
the potentials from Refs. \cite{Liu,Voter,Chen} require more delicate 
simulation technique. To check this point a new set of simulations for 
aluminum, copper, and gold described by the 
potentials from Refs. \cite{Liu,Voter,Chen} has been carried out. 
The same radius of $1.2$ nm was selected. In these new simulations
the starting geometry was the ordered configuration of the
five-shelled  nanowire which was obtained in simulations using the glue gold
potential \cite{GBSS}. The aluminum nanowire  
shown here in Fig.1 is also 
five-shelled,  but it is less ordered. In new simulations
the same procedure of annealing and quenching was applied.
The results are presented in Figs. 4, 5, and 6.
The aluminum nanowire in Fig. 4 consists of the fcc(111) layers. Copper and
gold nanowires in Figs. 5 and 6 also consist of the layers,
but there in the cross sections the atoms close to the cylindrical walls
follow circular paths.
Therefore, various filled and multi-shelled metallic nanowires are
stable within a long simulation time and a sensitive dependence of
the initial conditions was found.
Fluctuations in physical properties of metallic nanowires are studied
theoretically within the jellium model
\cite{Chaos}. The origin of these phenomena is 
the onset of chaos in classical and quantum billiard structures.
It should be mentioned that we did not find 
a such strong dependence on the initial configuration 
in gold nanowires modeled by the glue potential \cite{Goranka,GBSS}.
In these simulations gold shells form almost immediately after 
the simulation starts and regardless of an initial configuration.
The glue potential for gold, in comparison
with the potential of Voter and coworkers, gives a better description
of reconstruction  on gold surfaces \cite{Marco,Chen}. The subtle details of
inter-atomic interactions, that are responsible for surface reconstruction,
also give rise to internal reconstructuring of nanowires.
Therefore, it is possible that the 
multi-shelled structure (found for the glue
gold potential and here for the aluminum in Fig.1) is the most
frequent in fcc metallic nanowires.

In conclusion, it was shown that multi-shelled and filled metallic
nanowires exist  for several fcc metals.
The method of molecular dynamics simulation and 
the same kind of the embedded atom potentials for aluminum, copper, and gold
were used. 
The structures obtained strongly depend on the initial configuration.
In these simulations, no apparent correlation of the structure of nanowires
with the properties of metals was found. 

\acknowledgments
This work has been carried under the CRO-MZT project 
119206 - ``Dynamical Properties of Surfaces'' 
and the EC Research Action COST P3 - ``Simulation of Physical Phenomena
in Technological Applications''.

\clearpage

\begin{figure}
\caption{Infinite aluminum nanowire
with an average radius of $1.2$ $nm$, shown at $300$ K,  after $10^6$ 
time steps, and starting from an ideal fcc(111) structure:
(a) top view,
(b) side view.
The trajectory plots refer to a time span of $\sim 7$ $ps$
and include all atoms in the slice of the thickness of $4$ $nm$ 
along the wire axis.}
\label{fig1}
\end{figure}

\begin{figure}
\caption{
Copper nanowire simulated starting from the fcc(111) orientation:
(a) top view,
(b) side view.
Other details as in Fig. 1}
\label{fig2}
\end{figure}

\begin{figure}
\caption{
Gold nanowire simulated starting from the fcc(111) orientation:
(a) top view,
(b) side view.
Details as in Fig. 1}
\label{fig3}
\end{figure}

\begin{figure}
\caption{
Aluminum nanowire simulated starting from the five-shelled structure:
(a) top view,
(b) side view.
Details as in Fig. 1}
\label{fig4}
\end{figure}

\begin{figure}
\caption{
Copper nanowire simulated starting from the five-shelled structure:
(a) top view,
(b) side view.
Details as in Fig. 1}
\label{fig5}
\end{figure}

\begin{figure}
\caption{
Gold nanowire simulated starting from the five-shelled structure:
(a) top view,
(b) side view.
Details as in Fig. 1}
\label{fig6}
\end{figure}

\end{document}